Title:         Positive feedback and noise activate the stringent response regulator Rel in mycobacteria

Authors:       Kamakshi Sureka[1], Bhaswar Ghosh[2], Arunava Dasgupta[1*], Joyoti Basu[1], Manikuntala Kundu[1#] and Indrani Bose[2]

Address:       [1]Department of Chemistry and [2]Department of Physics, Bose Institute, 93/1 Acharya Prafulla Chandra Road, Kolkata 700009, India

Running title:    Mycobacterial Rel activation

*Present address: Department of Immunology, Max Planck Institute for Infection Biology, Chariteplatz 1, D10117 Berlin, Germany
#To whom correspondence should be addressed. E-mail: manikuntala@vsnl.net

**ABSTRACT**

Phenotypic heterogeneity in an isogenic, microbial population enables a subset of the population to persist under stress. In mycobacteria, stresses like nutrient and oxygen deprivation activate the stress response pathway involving the two-component system MprAB and the sigma factor, SigE. SigE in turn activates the expression of the stringent response regulator, *rel*. The enzyme polyphosphate kinase 1 (PPK1) regulates this pathway by synthesizing polyphosphate required for the activation of MprB. The precise manner in which only a subpopulation of bacterial cells develops persistence, remains unknown. Rel is required for mycobacterial persistence. Here we show that the distribution of *rel* expression levels in a growing population of mycobacteria is bimodal with two distinct peaks corresponding to low (L) and high (H) expression states, and further establish that a positive feedback loop involving the *mprAB* operon along with stochastic gene expression are responsible for the phenotypic heterogeneity. Combining single cell analysis by flow cytometry with theoretical modeling, we observe that during growth, noise-driven transitions take a subpopulation of cells from the L to the H state within a "window of opportunity" in time preceding the stationary phase. It is these cells which adapt to nutrient depletion in the stationary phase via the stringent response. We find evidence of hysteresis in the expression of *rel* in response to changing concentrations of PPK1. Hysteresis promotes robustness in the maintenance of the induced state. Our results provide, for the first time, evidence that bistability and stochastic gene expression could be important for the development of "heterogeneity with an advantage" in mycobacteria and suggest strategies for tackling tuberculosis like targeting transitions from the low to the high *rel* expression state.

**INTRODUCTION**

Tuberculosis, responsible for more than a million deaths every year, constitutes a major health concern worldwide. The effectiveness of the killer disease stems from the fact that the mycobacterial pathogen *M. tuberculosis*, the causative agent of the disease, is remarkably resilient against various physiological and environmental stresses including that induced by drugs. The "successful" containment of tubercular infection either through the host immune response or by the application of drugs is beset by the problem of persistence [1]. When subjected to stress, a fraction of the bacterial population survives through adaptive response in host tissues (lesions) over long periods of time [2]. These survivors are reactivated once the host immune system is compromised giving rise to a new bout of infection. The failure to tackle persistent infection is a major stumbling block in efforts to control tuberculosis. Understanding the phenomenon of persistence is therefore one of the priorities in tuberculosis drug development programs.

On tubercular infection, granulomas form in the host tissues enclosing the infected cells. Mycobacteria encounter an altered physical environment in the restricted space of granulomas with a paucity of life-sustaining constituents like nutrients, oxygen and iron [3,4]. The pathogens adapt to the stressed conditions and can survive over years in the so-called latent state. In vitro too, *M. tuberculosis* has been found to persist for years in the latent state characterized by the absence of active replication and metabolism [5]. Researchers have therefore attempted to develop models simulating conditions believed to be encountered by mycobacteria within the granuloma. One such model is the adaptation to nutrient-depleted stationary phase [6]. The processes leading to the slowdown of replicative and metabolic activity constitute the stringent response. In mycobacteria, the expression of *rel* initiates the stringent response which leads to persistence. The importance of Rel arises from the fact that it synthesizes the stringent response regulator ppGpp (guanosine tetraphosphate) [7]. Rel is essential for the long-term survival of *M. tuberculosis* under starvation [8] and for prolonged life of the bacilli in mice [9], suggesting that it is linked to persistence.

Persistence is a general phenomenon observed in microbial populations subjected to stress, e.g., when kept in fluctuating environments [10]. It is often linked to phenotypic heterogeneity arising in a genetically identical bacterial cell population with persisters forming a subset of the total cell population. Persistence against antibiotic treatment has been investigated in *E. coli* at the single cell level using microfluidic devices [10]. The experimental results suggest that stochastic transitions between normally growing cells and persister cells with reduced growth rates create phenotypic heterogeneity in bacterial populations even before the application of drugs. The preexisting heterogeneity may be advantageous for a bacterial population as it enables a fraction of the population to better cope with unfavorable circumstances. Phenotypic heterogeneity is often an outcome of gene expression dynamics involving positive feedback [11-13]. Two alternative scenarios have been proposed in this context. In one, positive feedback gives rise to multistability, i.e., the existence of more than one stable state for the same parameter values. The case most well-explored is that of bistability which gives rise to two stable expression states: low (L) and high (H). A transition from the L to the H state occurs when the magnitude of an appropriate parameter, say, inducer amount, exceeds a threshold value. Gene expression noise, i.e., fluctuations in expression levels also play a significant role in driving the transitions between the stable states. In this scenario, phenotypic heterogeneity implies the coexistence of two distinct subpopulations in L and H states. A recent experiment [14] on competence development in *B. subtilis* established the role of noise in driving the phenotypic transitions from the non-cometent to the competent state. The network of regulatory interactions controlling competence development contains a core module, an autostimulatory positive feedback loop in which ComK proteins, synthesized from the *comK* gene, promote their own production. The non-competent and competent states correspond to low and high ComK levels. In the alternative scenario leading to positive feedback-based phenotypic heterogeneity, there is one stable expression state, mostly, the low expression state. The high expression state is unstable so that noise-driven activation to this state is followed by a relaxation to the low expression state. Competence development in *B. subtilis* has also been analysed in this alternative scenario [15] with the core circuit containing both a positive and a negative feedback loop. Evidence for a single stable expression state and transient activation to an alternative expression state has been obtained in some other cases [16].

Recent experiments provide knowledge of the stress signaling pathway in mycobacteria linking polyphosphate (poly –P), the two-component response regulator MprAB, the alternate sigma factor SigE and Rel [17]. Figure 1 shows a sketch of the important components of the pathway. *mprA* and *mprB* encode the histidine kinase sensor MprB and its partner the cytoplasmic response regulator MprA respectively. The protein pair responds to environmental stimuli by initiating adaptive transcriptional programs. Polyphosphate kinase 1 (PPK1) catalyses the synthesis of polyphosphate (poly P) which is a linear polymer composed of several orthophosphate residues. MprB autophosphorylates itself with poly P serving as the phosphate donor [17].The phosphorylated MprB-P phosphorylates MprA via phosphotransfer reactions. There is also evidence that MprB-P functions as a MprA-P (phosphorylated MprA) phosphatase [18]. MprA-P binds the promoter of the *mprAB* operon [19] to initiate transcription. A positive feedback loop is functional in the network of interactions as the production of MprA brings about further MprA synthesis. The *mprAB* operon has a basal level of gene expression [18,19] irrespective of the operation of the positive feedback loop. Once the *mprAB* operon is activated, MprA-P regulates the transcription of the alternate sigma factor gene Sig E, which in turn controls the transcription of *rel*.

We hypothesized that positive feedback and stochastic gene expression likely play a role in the development of persistence in mycobacteria. To test this hypothesis, we studied the dynamics of *rel* expression in mycobacteria grown up to stationary phase, a model where nutrient depletion serves as the source of stress. We combined mathematical modeling with experiments to establish that bistability along with gene expression noise give rise to phenotypic heterogeneity in terms of the expression of *rel*. Our results indicate that noise influences cell fate under stringency in mycobacteria. By regulated expression of PPK1 from a tetracycline-inducible promoter in a *ppk1* knockout mutant, we further demonstrated hysteresis in *rel* expression.

**RESULTS**

**Mathematical modeling of the *mprAB*-driven stress signaling pathway**

The positive autoregulation of the master regulator *mprAB* (Figure 1) draws attention to the recently appreciated role of positive feedback loops in certain bacteria giving rise to bistability [11]. We constructed a mathematical model to study the dynamics of the above signaling pathway. The reaction scheme and the calculations are reported in the Supporting Information (SI). The theoretical study predicted the existence of bistability (Figures S1A-C) and provided the motivation to test the prediction in an actual experiment. In a deterministic scenario, all the cells in a population are in the same steady state if exposed to the same environment. To explain heterogeneity in a genetically identical cell population, a stochastic (probabilistic) description of cellular events is necessary. Gene expression involves biochemical events which are inherently probabilistic [20,21]. The uncertainty introduces fluctuations (noise) around mean expression levels so that the single protein level of the deterministic model broadens into a distribution of levels. In the case of bistable gene expression, the distribution of protein levels in a population of cells is bimodal with two distinct peaks.

**Bimodal *rel* Expression in *M. smegmatis***

Considering that persistent mycobacteria in lung lesions are nutritionally starved [8], the issue of persistence has been addressed in terms of how mycobacteria adapt to stress in the form of nutrient-depletion. The stringent response is a broad transcriptional program in prokaryotes that becomes operative under stress in order to enable survival of at least a fraction of the cell population. Key elements of the stringent response and the ability to survive over long periods of time under stress are shared between *M. tuberculosis* and *M. smegmatis* [22]. The protein Rel plays a central role in this stress response program. We therefore chose to study the dynamics of *rel* transcription in individual cells of *M. smegmatis* grown in nutrient medium up to stationary phase, with nutrient depletion serving as the source of stress. The experimental signature of bistability lies in the coexistence of two subpopulations. We employed flow cytometry to monitor the dynamics of green fluorescent protein (GFP) expression in *M. smegmatis* harboring the *rel* promoter fused to *gfp* [17] as a function of time. Initially, the fraction of cells with "low" GFP expression (L) is high (Figure 2). The fraction of cell population in the "high" GFP expression state (H) increases as a function of time due to transitions from the L to the H state. The coexistence of H and L cells in the population was also observed by fluorescence microscopy (Figure S2). The distribution of GFP-expressing cells (Figure 2) is bimodal indicating the existence of two distinct subpopulations. In the stationary phase [i.e. around 36 h (Figure S3)], the majority of cells (~ 70 %) is in the H state. We term the GFP level in the L state as the basal level. Plots of the mean basal level (fluorescence intensity) (Figure 3A) and the fractions of cells in the H state (Figure 3B) against time could be fitted with sigmoids. The presence of two distinct subpopulations in the experiment confirmed the theoretical prediction of bistability. If clumping of mycobacterial cells and cell-to-cell variation of plasmid copy number were responsible for the observed bimodal fluorescence intensity distribution in Figure 2, bimodality would also likely be observed in the case of GFP expression being driven by a constitutve promoter. As a control, we therefore chose to analyze GFP expression driven by the constitutive *hsp60* promoter as a function of time. A single bright population was observed at different times of growth (Figure S4). The unimodal, rather than bimodal distribution ruled out the possibility that the bimodality of *rel* promoter-driven GFP expression was artifactual. Our earlier observations have shown that *mprAB* expression increases with time of growth [17] and that *mprAB* regulates *rel* expression through its ability to drive the transcription of *sigE*. In order to establish that positive feedback is responsible for the bimodal distribution of GFP expression, we analyzed the time course of *rel*-driven GFP expression in a *M. smegmatis* knock-out mutant in which the chromosomal copy of *sigE* was disrupted (i.e. a situation in which the positive feedback loop activating *mprAB* expression, was decoupled from *rel* expression). In this strain, only a negligible fraction of the cell population was in the H state at any phase of growth (Figure 4A), supporting the view that positive feedback is essential for the bimodal distribution of GFP expression. The promoter of *rel* contains binding sites for SigA and SigE (Figure 4B). Mutation of the SigE binding site resulted in a negligible fraction of the cells being present in the H state (Figure 4C). On the contrary, we observed bimodality (Figure 4D) when the SigA binding site was mutated. Taken together, these results confirmed the role of MprAB/SigE-mediated positive feedback in bimodal *rel* expression in mycobacteria.

**Gene Expression Noise Drives Stochastic Transitions**

We next probed the role of gene expression noise in effecting the switch from the L to the H state. A bistable system in the L state can undergo a transition to the H state if the basal level crosses a threshold value [SI]. We analyzed the rate of transition from the L to the H state as a function of time (Figure 3C). The transition rate initially increases with time, reaches a maximum (at around 20 h) and falls to zero before the cell population enters the stationary phase, suggesting that transition to the H state takes place within a limited "window of opportunity" in time. The inset of Figure 3C shows how the basal expression rate changes as a function of time. Flow cytometry measurements provided an estimate of the coefficient of variation (CV), a quantifier of noise (CV = standard deviation/mean), associated with the gene expression level. A plot of CV of the basal expression level of *rel* as a function of time (Figure 3C) showed a similar variation as the transition rate, with the maxima in both the curves occurring around the same time. A steep increase in the CV of the basal expression level is accompanied by a steep increase in the rate of transition to the high GFP expression state. The basal expression level has a low value and a comparatively slower increase during the same time period. Since CV is the ratio of standard deviation and mean, it appears that increasing basal expression fluctuations are predominantly responsible for the increasing rate of transition to the high expression state. To achieve a specific fraction of cells in the H state, the basal GFP level in these cells should cross a threshold value [SI]. This is possible in two ways: with low mean basal level and a large standard deviation or a high mean basal level with a low standard deviation. Figure 3 shows that the first strategy is adopted in this case. The choice could be dictated by the metabolic cost of sustaining a high mean basal level in the cells.

**Hysteresis in *rel* Expression**

In order to obtain further definitive evidence of bistability in *rel* expression, we attempted to detect hysteresis by analyzing the "history-dependent" response of the system. To demonstrate hysteresis in our system, we took advantage of the *ppk1* knockout mutant (PPK-KO) of *M. smegmatis* [17] which did not show *rel* expression. We reasoned that regulated expression of PPK1 would serve to tune the positive feedback by virtue of the ability of the enzyme to synthesize poly P, the phosphate donor for phosphorylation of MprA. We therefore introduced the *ppk1* gene under the control of the *tet* promoter [23] in PPK-KO harboring the *rel-gfp* plasmid, and confirmed tetracycline-inducible *rel* expression by fluorescence microscopy (Figure 5). The generic inducer-response curve predicted theoretically (Figure S1C) provided evidence of bistability. For the experimental analysis of hysteresis, we grew PPK-KO carrying the tetracycline-inducible [23] *ppk1* and *rel-gfp* in medium with increasing concentrations of tetracycline (inducer) and analyzed the distribution of cells expressing GFP by flow cytometry in the stationary phase (steady state) (Figure 6A, going up). Similarly, we obtained the branch (Figure 6A, coming down) for decreasing concentrations of tetracycline with the mean GFP expression measured in the stationary phase. The standard error of each data point is not shown because of its small magnitude. We observed two distinct alternative states within a particular range of inducer concentrations depending on how the concentrations were changed. The distinct branches in Figure 6 confirm the existence of hysteresis supporting our prediction of bistability in the system. We analyzed the levels of PPK1 by Western blotting (Figure S5A,B) as well as levels of poly P (Figure S5C) in the "going up" and "coming down" experiments. Hysteresis was not observed in relation to PPK1 or poly P. Thus, hysteresis in *rel* expression levels could not be attributed to differences in levels of PPK1 or poly P in the two sets of experiments. We confirmed the critical role of PPK1 in *rel* expression by carrying out the same experiments (as a control) in PPK-KO harboring a tetracycline-inducible kinase-dead mutant of *ppk1*. No induction of *rel* expression was observed when bacteria were grown in the presence of tetracycline (data not shown) confirming the role of the enzymatic activity of PPK1 in the induction of *rel* expression. The set of GFP distributions as a function of the inducer concentration are shown in Figure 7. Stochastic simulation of the dynamical model (Figure 1) based on the well-known Gillespie algorithm [24] (SI) reproduces the qualitative features of the distributions shown in Figure 2 (Figure S6) and the plots in Figure 3C (Figure S7).

**DISCUSSION**

The phenomenon of persistence has been investigated in a host of microorganisms but the molecular mechanism which activates the switch to persistence in a cell is little understood in most cases. Persistence can develop in response to a variety of stresses some of which, like nutrient and oxygen deprivation, bring about the stringent response in bacterial systems. Rel is widely acknowledged as a key player enabling mycobacteria to persist under conditions of stress. In this paper, we identified the triggering mechanism of the switch to high *rel* expression in a growing population of mycobacteria with nutrient depletion serving as the source of stress. Summarising our present findings, we observe that *rel* expression is bistable and a combination of positive feedback along with gene expression noise select a subset of cells to be in the high *rel* expression state. Our results suggest that there is a temporal window during which fluctuations in the basal expression level drive the transitions from the low (L) to the high (H) expression state. Similar experimental observations have been made on competence development in *B. subtilis* [14,25] in which phenotypic heterogeneity is linked to the coexistence of competent and non-competent cells.

Recent studies have pinpointed the contributions of positive feedback and gene expression noise towards creating phenotypic heterogeneity in *B. subtilis* [11,12,14,15, 25,26]. Two different mechanisms have been proposed to explain competence development in a fraction of cells: the case of bistability with two stable gene expression states [11,26] and excitability [15] with one stable gene expression state. In our experiments on mycobacteria, we find evidence of bistability which gives rise to a bimodal distribution in *rel* expression in the stationary phase. We also find experimental evidence of another important feature of bistability, namely, hysteresis. Hysteresis is known to promote robustness in the maintenance of the induced state [13,27]. In the case of excitability, the induced state is transient. In a bistable system, gene expression noise can give rise to transitions between the two stable states. In our experiments, stochastic switching is predominantly from the low to the high expression state with the latter being quite stable.

Cells in the H state in the mycobacterial population are generated in the exponential phase itself hinting at pre-existing heterogeneity in the cell populations. Such heterogeneity has been suggested [10] to be a strategy of "hedging the bet" adopted by microorganisms to be better prepared for coping with changing environmental conditions. Consistent with our view that positive feedback at the level of *mprAB* generates phenotypic heterogeneity, we observed that *sigE*-GFP expression is also bimodal (Figure S8).

One of our major findings relates to the observation (Figure 3C) that the rate of transition from the low to the high expression state is maximum close to the time point at which the coefficient of variation (CV) of the basal expression level attains the largest value. This result appears to suggest that transitions to the high expression state are aided by basal level noise. Mamaar et al. [14] studied competence development in *B. subtilis* and suggested that intrinsic noise associated with *comK* mRNA production and degradation are possibly major sources of variation in ComK protein levels. They tested the hypothesis by experimentally reducing the noise. This was achieved through increasing the rate of *comK* gene transcription and simultaneously reducing the translation rate by an equal amount so that the mean protein level remains the same. In this experiment, the reduction in noise led to fewer transitions to the competent state since basal level fluctuations required to trigger the positive feedback loop were lower in magnitude. A similar conclusion was reached in an earlier theoretical study [26] in which the amount of noise was changed by varying the rate constants for gene activation and deactivation keeping the mean expression level the same. Recent modeling and experimental results appear to reinforce the view that the combined effect of positive feedback and noise provide a universal mechanism for generating phenotypic heterogeneity in cell populations [11,12,14-16,25,26,28].

As has been recently pointed out [29], the network of regulatory interactions controlling the stringent response likely differs in detail from that in *E. coli*. Elevated levels of (p)ppGpp favor the accumulation of poly P in *E. coli* [30]. On the other hand,  in mycobacteria, poly P is essential for activating *rel* expression at least under a certain set of conditions [17] and subsequent biosynthesis of (p)ppGpp. Though the importance of poly P and the regulatory link between *sigE* and *rel* is now well established [17], a number of questions remain unanswered. These include how *ppk1* expression is regulated, whether *mprAB* preferentially utilizes poly P and whether only *ppk1* transcription is responsible for poly P accumulation [29]. Considering the role of *rel* in persistence, it would also be of great interest to analyze whether the high-*rel* population is antibiotic-tolerant and represent persistors. While the present study does not address these questions, our characterization of the high *rel* state and a systems level analysis of how mycobacteria attain this state offers important clues regarding possible drug targets. One aim of drug therapy could be the inhibition of the switching transition from the low to the high *rel* expression state thus reducing the chances of mycobacterial survival via the stringent response.

## MATERIALS AND METHODS

### Strains

*M. smegmatis* mc$^2$155 [31] was grown in Middle Brook 7H9 medium supplemented with 2 % glucose and 0.05% Tween 20.

### Construction of plasmids for fluorescence measurements

The *rel* promoter (−610 to +1 relative to the translational start site) fused to *gfp* in pFPV27 has been described [17]. Mutants of the *rel* promoter were generated by overlap extension PCR. The primers used are depicted in Supplementary Table S1. The initial rounds of PCR were carried out using primer pairs a and b, and c and d. The products of each PCR were purified and used as templates for the second round of PCR using the primers a and d. The resulting plasmid carrying the wild type or mutant *rel* promoter was electroporated into wild-type *M. smegmatis* mc$^2$155, or the isogenic strain inactivated in the *ppk1* gene (PPK-KO) [17], or PPK-KO harboring the *ppk1* gene under the control of a tetracycline-inducible promoter or in the *sigE*-inactivated  (*sigE* KO) strain RH243, a gift from Prof. R.N. Husson [32].  As a positive control the constitutive *hsp* promoter was excised from pCKB115 [33] and  fused to *gfp* in  the replicative promoter-less gfp vector pFPV27 [34]

The *sigE*  promoter[32] was amplified from the genomic DNA of *M. smegmatis* using the sense and antisense primers

5'-TAA**GGTACC**GCGGGATACAGTTCCCCATGC-3' [KpnI site in bold]

5'-TA**GGATCC**TTCAACATGCTAAACGCATGT-3' [BamHI site in bold]

respectively and cloned into the promoter-less replicative gfp vector pFPV27 [33] using asymmetric KpnI and BamHI sites. The resulting plasmid was electroporated into wild-type *M. smegmatis* mc$^2$155, the isogenic PPK-KO and the *tet-ppk1*-complemented strain.

**Expression of *ppk1* under a tetracycline-inducible promoter**

Expression of *ppk1* under a tetracycline-inducible promoter was achieved by constructing an integrating vector containing a hygromycin-resistance cassette along with a positive-selection L5 integrase cassette. *ppk1* was first amplified from genomic DNA using the sense and antisense primers 5'-TAA**CATATG**ATAAGCAATGATCGCAAG-3' [NdeI site in bold] and 5'-TTA**AAGCTT**CAGGGGCTGCGGTGCCGTT-3' [HindIII site in bold] respectively and cloned between same sites of pMIND [23] under the control of the *tet* promoter. The *ppk1* gene along with the *tet* promoter was excised by digesting with KpnI and HindIII and cloned between the same sites in pUC19 to generate pTC-PPK. A 3.7 kb Hyg-integrase cassette from pUC-HY-INT [35] was cloned at the HindIII site of pTC-PPK to generate the integrating construct pTC-PPKINT which was electroporated into PPK-KO. Hygromycin-resistant colonies were selected and analyzed by PCR to confirm the presence an integrated copy of the wild-type *ppk1*. The kinase-dead *ppk1* carrying mutations of two histidine residues (required for autophosphorylation) to alanine [17] [*ppk1-KD*] was amplified by PCR from a suitable construct essentially as described by Sureka et al. [17] using the above primers and integrated into *M. smegmatis* mc$^2$155 as described above.

**FACS analysis**

*M. smegmatis* cells were grown for different periods of time, harvested, washed twice and resuspended in PBS. Fluorescent bacteria were analyzed on a FACS Caliber (BD Biosciences) flow cytometer. All data modes were set to log, and the forward scatter detector level was set to E-1. The fluorescence intensity of 20,000 ungated events was measured in FL-1 at a detector sensitivity of 780. A side scatter threshold was set to gate out inappropriate heterogeneities. It is difficult to analyze single mycobacteria by FACS because of their tendency to clump. Careful gating procedures were adopted to minimize the effect of clumping. The data files were analyzed using CellQuest Pro (BD Biosciences) and WINMIDI. The flow cytometry data is represented in histogram plots where the x-axis is a measure of fluorescence  intensity and the y-axis represents the number of events.

**Fluorescence microscopy**

Cells were fixed by incubation for 15 minutes at room temperature followed by  45 minutes on ice in 2.5% (v/v) paraformaldehyde, 0.04% (v/v) glutaraldehyde, 30 mM sodium phosphate (pH 7.5). The cells were transferred to slides, washed with PBS, air-dried, dipped in methanol (-20°C) for 5 min and then in acetone (-20°C) for 30 s and allowed to dry. Brightfield and fluorescence images were taken using a 100x oil immersion objective on a Zeiss Axioimager A1 microscope equipped with a cooled CCD camera.

**ACKNOWLEDGEMENTS**

We thank Dr. R.N. Husson for the gift of the *sigE* mutant of *M. smegmatis*, and Drs. Richard Stokes, Brian Robertson and Lalita Ramakrishnan for the plasmids pUC-HY-INT, pMIND and pFPV27  respectively. B.G. was supported by the Council of Scientific and Industrial Research, Government of India. This work was supported in part  by a grant from the Department of Biotechnology, Government of India.

**Author contributions**

K.S. and B.G. contributed equally to the study.

# LEGENDS TO FIGURES

**Fig. 1**. **Schematic diagram of MprAB-SigE-mediated *rel* transcription**. MprB-P and MprA-P denote phosphorylated forms of MprB and A respectively. Poly P serves as the phosphate donor in the conversion of MprB to MprB-P.

**Fig. 2**. **Time course of *rel*-GFP expression**. *M. smegmatis* harboring a *rel*-GFP construct was grown for different periods of time (indicated in h) and the *rel* promoter-driven expression of GFP was monitored by flow cytometry. With time, there is a gradual transition of the cells from the L (low) to the H (high) state.

**Fig. 3**. **Analysis of the time course of *rel* expression**. (A) Basal (mean fluorescence intensity of cells present in the L state) *rel* expression and (B) the percentage of cells in the high state were obtained from flow cytometry data and plotted against time. The data in both the cases can be fitted with sigmoids. (C) Transition rate from the L to the H state (solid curve) and the low expression state CV (solid squares represent data points) versus time. The inset shows the basal expression rate versus time.

**Fig. 4.** Time course of *rel*-GFP expression in *M. smegmatis*. A. schematic diagram of *rel* promoter. B. Time course of *rel*-GFP expression in a *M. smegmatis* mutant in which *sigE* was disrupted. Time course of GFP expression driven by *rel* promoter mutated at the SigE (C) or the SigA (D) binding site. The time points (h) in A, C and D are indicated in the histograms.

**Fig. 5.** Fluorescence microscopy of tetracycline (tet)-inducible GFP expression. PPK-KO harboring *rel-gfp* and *tet*-inducible *ppk1* was left uninduced (A-C) or induced in the presence of tet (D-F). A, D: phase contrast; B, E: fluorescence micrographs; C, F: merge.

**Fig. 6**. Hysteresis in *rel* expression. (A) Filled triangles and squares represent the experimental data on expression of *rel* with increasing and decreasing (denoted by arrows) concentrations of inducer (tetracycline) respectively. (B) Hysteresis plot obtained through stochastic simulation based on the Gillespie algorithm. At a particular range of concentrations, we observed two distinct alternative steady states of rel expression depending on whether inducer concentration was going up or coming down. Figure 4C shows the GFP distributions in the stationary phase for two sets of experiments, one in which the inducer concentration is increased from a low to a specific value (indicated as "Low" in black) and the other in which the same value is reached by decreasing the inducer concentration from a high value (indicated as "High" in red). The first and third panels in which the specific inducer concentration is respectively low and high correspond to regions of monostability. The distributions with different histories, i.e., initial conditions more or less coincide. The middle panel describes a region of bistability. The distinct distributions for different histories indicate persistent memory (see SI).

**Fig. 7.** GFP distributions as a function of tetracycline inducer concentration. The inducer activates the transcription of the *ppk1* gene. The concentrations of tetracycline (nM) are indicated in the histograms.

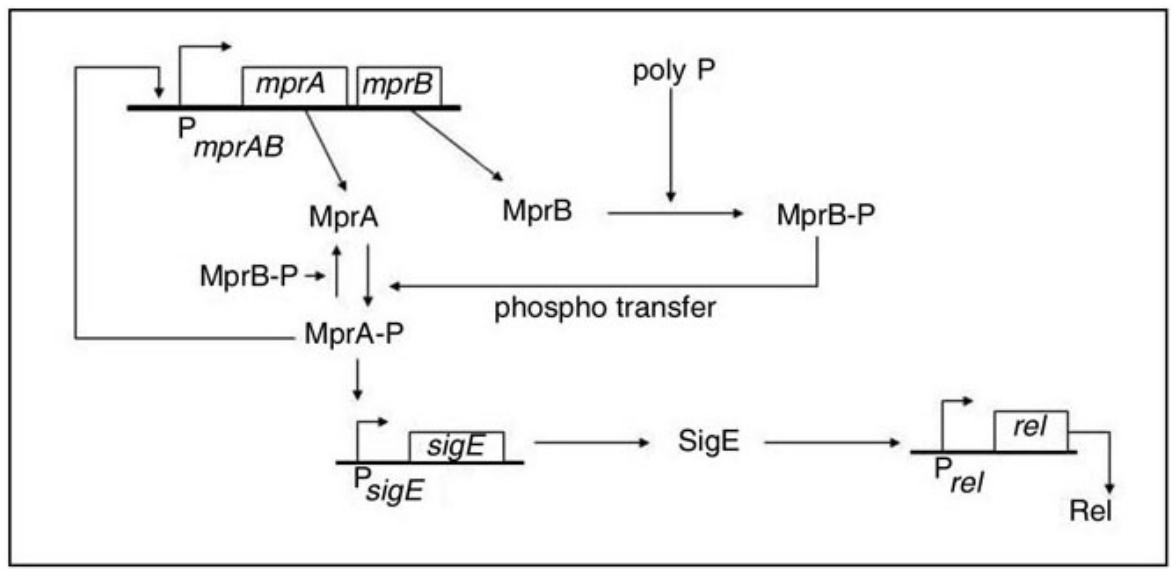

Fig. 1

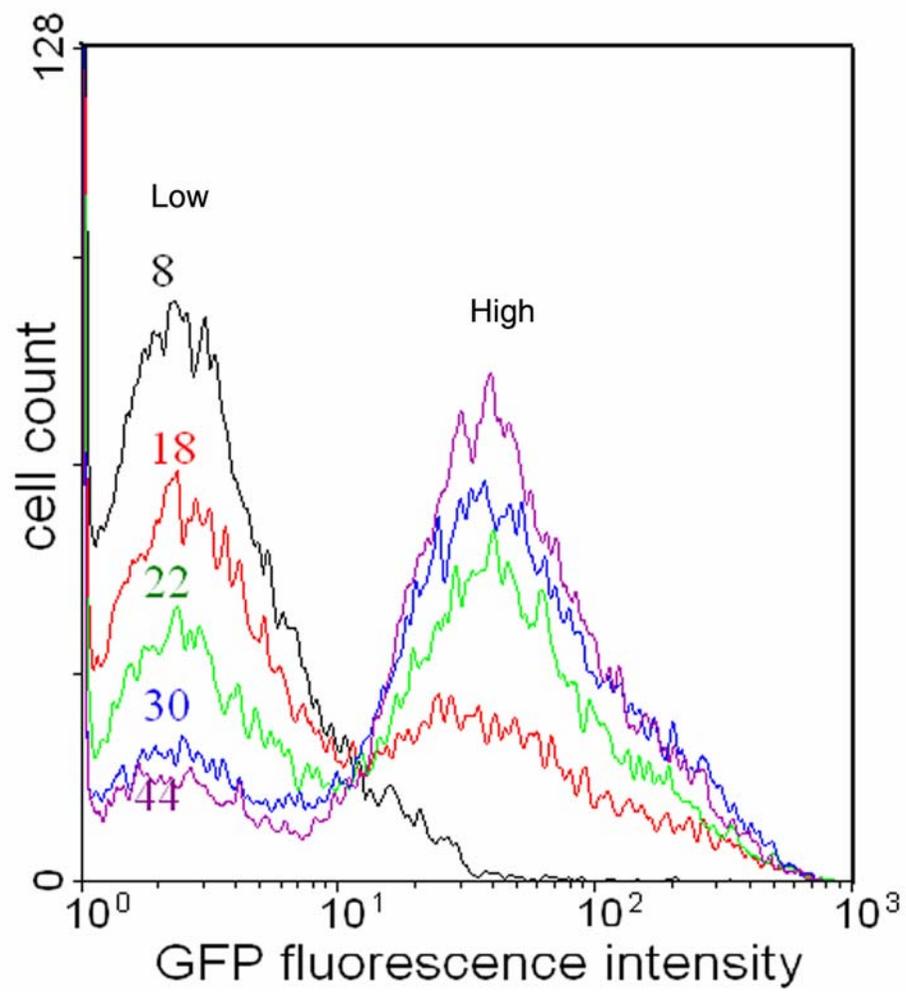

Fig. 2

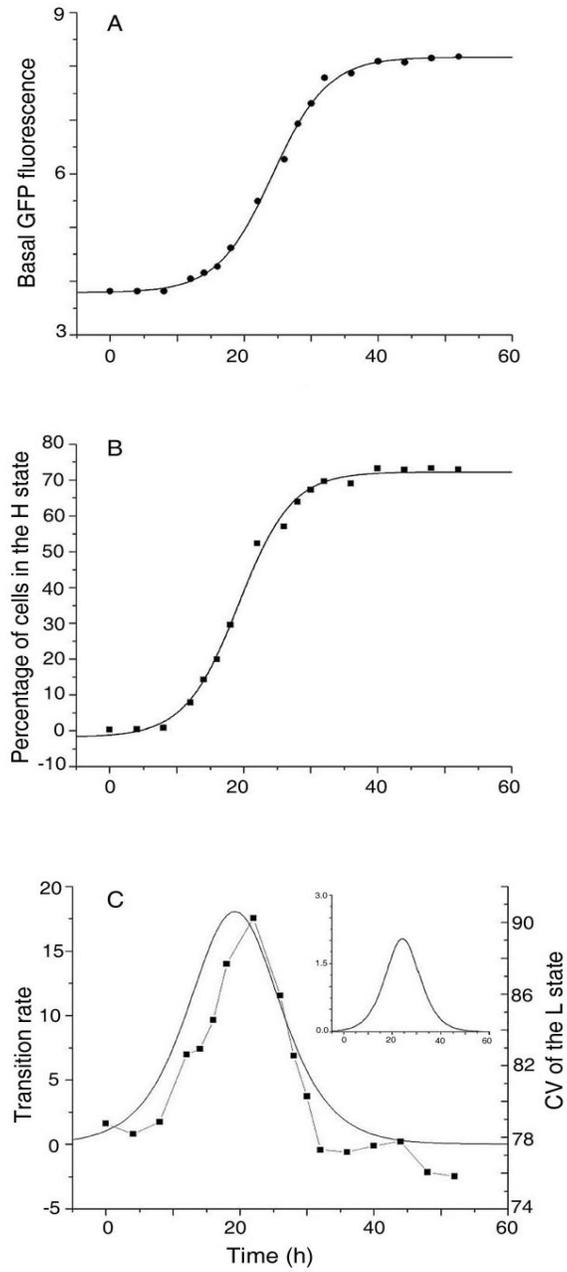

Fig. 3

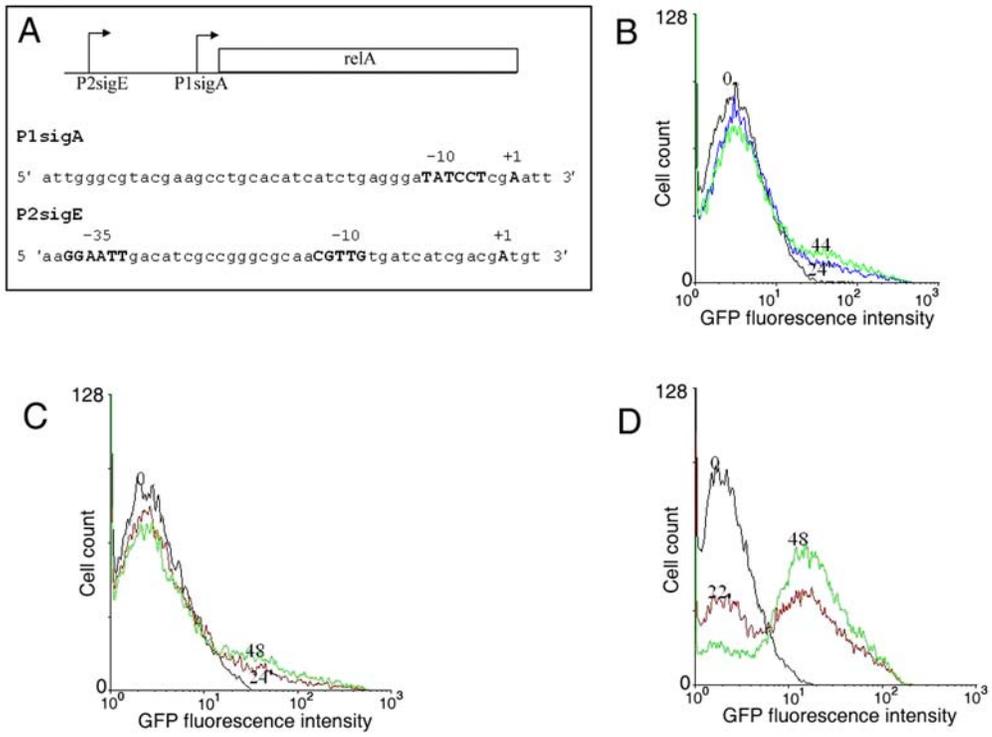

Fig. 4.

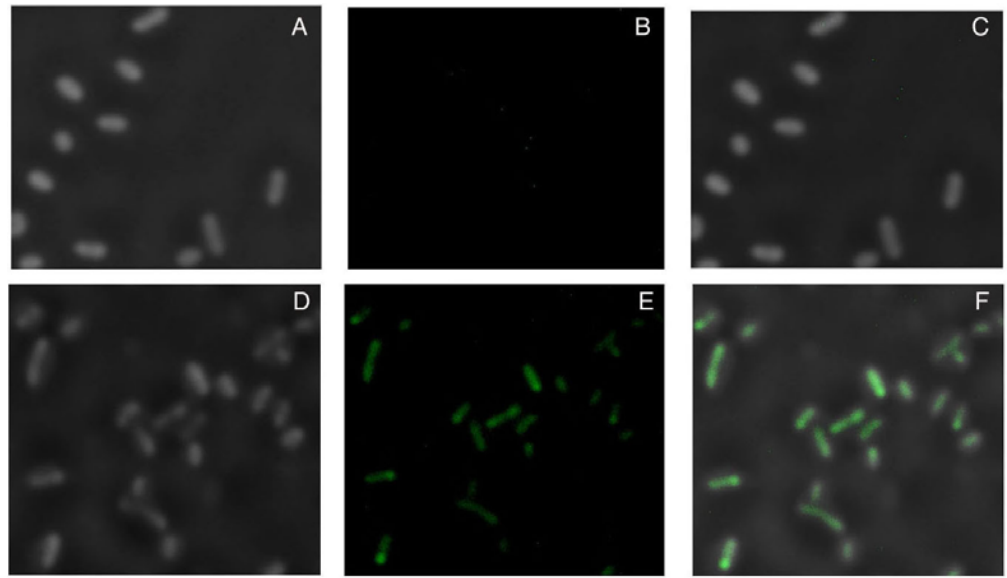

Fig. 5

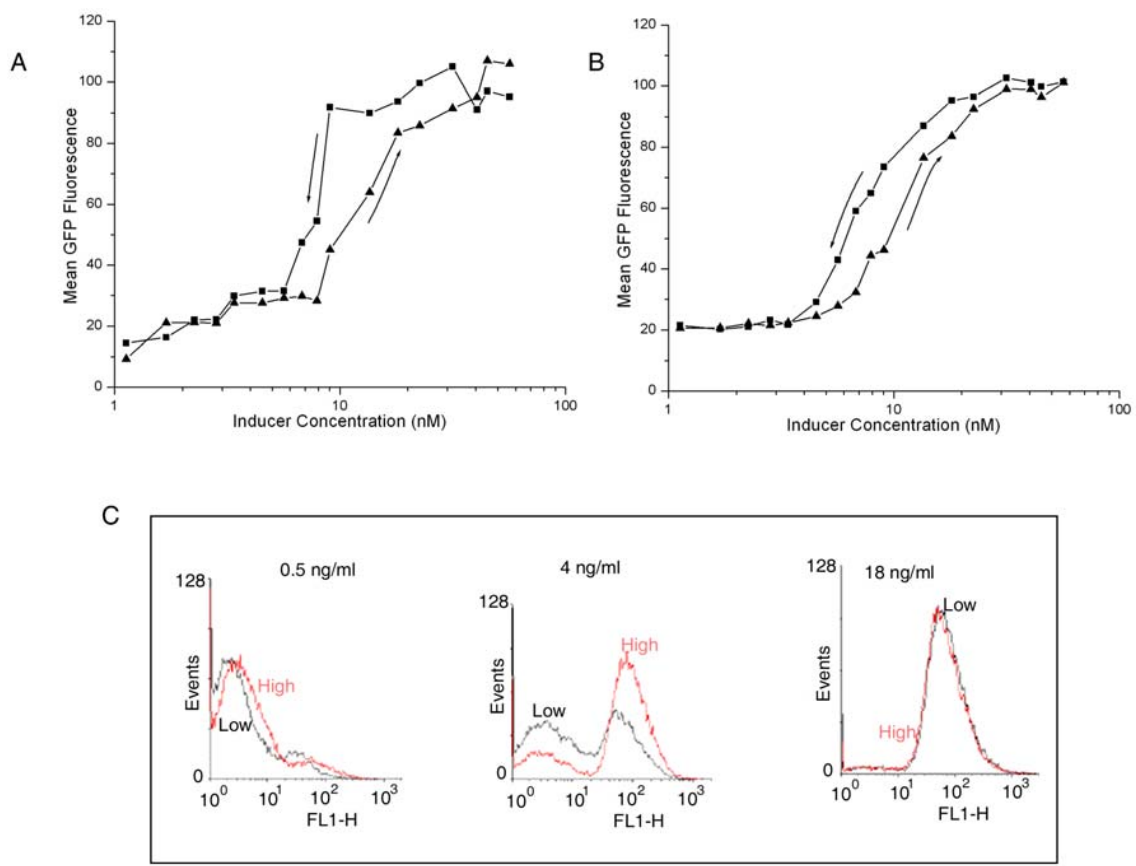

Fig. 6

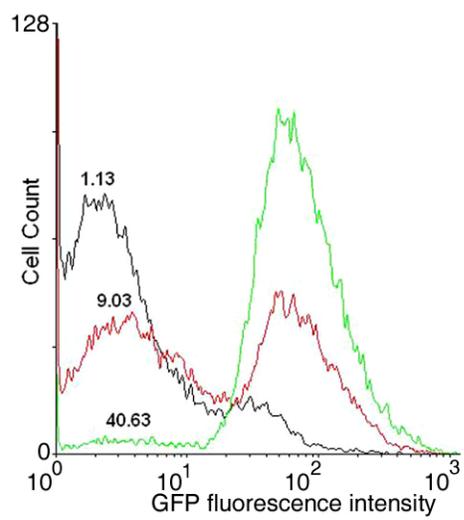

Fig. 7